\documentclass[conference]{IEEEtran}
\IEEEoverridecommandlockouts 
\usepackage{cite}
\usepackage{amsmath,amssymb,amsfonts}
\usepackage{algorithmic}
\usepackage{graphicx}
\usepackage{textcomp}
\usepackage{xcolor}
\usepackage{hyperref}
\usepackage{subcaption}

\def\BibTeX{{\rm B\kern-.05em{\sc i\kern-.025em b}\kern-.08em
    T\kern-.1667em\lower.7ex\hbox{E}\kern-.125emX}}
\begin{document}

\title{LIN-MM: Multiplexed Message Authentication Code for Local Interconnect Network message authentication in road vehicles}

\author{\IEEEauthorblockN{Franco Oberti\textsuperscript{1,2}\thanks{Authors contacts: \{franco.oberti, alessandro.savino, ernesto.sanchez, stefano.dicarlo\}@polito.it and filippo.parisi@punchtorino.com}, Ernesto Sanchez\textsuperscript{1}, Alessandro Savino\textsuperscript{1}, Filippo Parisi\textsuperscript{2},Mirco Brero\textsuperscript{2}, and Stefano Di Carlo\textsuperscript{1}}
\IEEEauthorblockA{\textsuperscript{1}\textit{Control and Computer Engineering Department,} Politecnico di Torino,
Torino, Italy \\
\textsuperscript{2}\textit{PUNCH Softronix S.r.l.}, Torino, Italy}
}

\maketitle

\begin{abstract}
The automotive market is profitable for cyberattacks with the constant shift toward interconnected vehicles. Electronic Control Units (ECUs) installed on cars often operate in a critical and hostile environment. Hence, both carmakers and governments have supported initiatives to mitigate risks and threats belonging to the automotive domain. 
The Local Interconnect Network (LIN) is one of the most used communication protocols in the automotive field. Today’s LIN buses have just a few light security mechanisms to assure integrity through Message Authentication Codes (MAC). However, several limitations with strong constraints make applying those techniques to LIN networks challenging, leaving several vehicles still unprotected. This paper presents LIN Multiplexed MAC (LIN-MM), a new approach for exploiting signal modulation to multiplex MAC data with standard LIN communication. LIN-MM allows for transmitting MAC payloads, maintaining full-back compatibility with all versions of the standard LIN protocol. 
\end{abstract}


\section{Introduction}
\label{sec:introduction}
Nowadays, the automotive field is turning into a profitable domain for attackers\cite{CySTrend}. Vehicles and related systems are operating in an enduring inimical environment\cite{8537180}. Therefore, carmakers and governments are becoming increasingly sensitive about road vehicle cyber-attacks.

The United Nations Economic Commission for Europe (UNECE)\cite{UNECE} is promoting explicit legislation on road vehicles cybersecurity. In particular, the existing Work of Parties on the Adoption of Harmonized Technical United Nations Regulations for Wheeled Vehicles (WP29) \cite{WP29} endorsed two new regulation bullets: UN Regulation No. 155 (UNR155) \cite{unece-155-2021} and UN Regulation No. 156 (UNR 156)\cite{unece-156-2021}. The first focuses on the vehicle's cybersecurity management system (CSMS), while the latter introduces the requirement of building a software update management architecture.
The ISO/SAE 21434:2021 Road vehicles Cybersecurity engineering requirements\cite{iso21434} is a new standard introduced by the automotive industry to secure all supply chain developments in compliance with UNR155 and UNR156 regulations.

Henceforth, every road vehicle, both permanently and seamlessly connected, must satisfy UNR155 and UNR156 regulations. Motorcycles and agriculture vehicles are unique temporary exceptions. Cybersecurity conformity is crucial because it consents access to vehicle markets under UNECE statutes. Each violation of these security directives forbids selling with a massive loss of money. With this upcoming scenario, the entire automotive industry has increased the effort to apply security in its products and improve research activities. 

The most widespread security threats in the automotive domain exploit communication channels among modules \cite{9525579,oberti2021taurum,9729479}. Several of these attacks exploit vulnerabilities of the primary automotive communication protocol, i.e., the Controller Area Network (CAN) protocol \cite{iso118983}. Therefore, the current literature proposes several solid secure mechanisms to protect CAN networks \cite{6542519} and make them resilient to attacks. However, other protocols, such as the Local Interconnect Network (LIN), are gaining importance in the automotive domain \cite{iso17987}. 

Four European carmakers (i.e., BMW, Volkswagen Group, Volvo Cars, Mercedes-Benz) formed the LIN consortium in the late 1990s. At the same time, both Volcano Automotive Group and Motorola provided consulting services, especially in the networking and hardware fields. The first LIN version 1.1 was deployed in 1999. A solid requirements document was available only in 2002, with version 1.3. One year later, version  2.0 expanded the diagnostic features. The latest known release of the protocol is version 2.2A \cite{iso17987} released in 2010. In 2016, the CAN in Automation (CiA) organization made the LIN protocol a standard (ISO 17987: 2016). Despite being of age, the LIN protocol is being spread in automotive applications following a positive trend of growth: starting from 200 million nodes in 2010 to reaching 700 million in 2020. Despite several studies confirming critical vulnerabilities associated with this protocol, \cite{9628576,hackLIN,DENG2017131,MARTINEZCRUZ20211}, the research community has been scarcely active in proposing efficient countermeasures. 

This work introduces LIN Multiplexed MAC (LIN-MM), a novel technique to multiplex a data digest of a LIN payload with the payload itself. The computed digest includes both a Message Authentication Code (MAC), and a Message Integrity Code (MIC)\cite{MAC-MIC} to guarantee the integrity and authenticity of transmitted LIN data frames. LIN-MM is designed to avoid modifying the original LIN data frame, thus providing full back compatibility with the standard protocol. Moreover, by multiplexing both data and authentication codes, robust optimization in the digest validation is provided, reducing the processing time.

The paper has the following organization: \autoref{sec:LIN-overview} introduces the basic organization of a state-of-the-art vehicle LIN network, while \autoref{sec:attack} describes the considered attack model on the LIN vector. Section \ref{sec:LINMM} describes the LIN-MM architecture, and \autoref{sec:results} provides experimental results and validation of the proposed solution. Eventually,  \autoref{sec:conclusions} summarizes the main contributions and concludes the paper.
\section{LIN Protocol overview}
\label{sec:LIN-overview}
Modern automotive applications embed more than one hundred Electronic Control Units (ECUs). ECUs are connected to sensors and actuators to manage and control all vehicle subsystems \cite{albert2004comparison}. Many events in this complex network are being processed in real-time.

The Local Interconnected Network (LIN) and Controller Area Network (CAN) are the main communication protocols used on standard vehicle networks. The CAN provides high-speed communication with solid reliability. The LIN serves domains where high performance and reliability are not primary targets. Therefore, the LIN protocol is a viable solution to building up a low-cost vehicle communication network.

\begin{figure}[htb]
   \centering
    \includegraphics[width=\columnwidth]{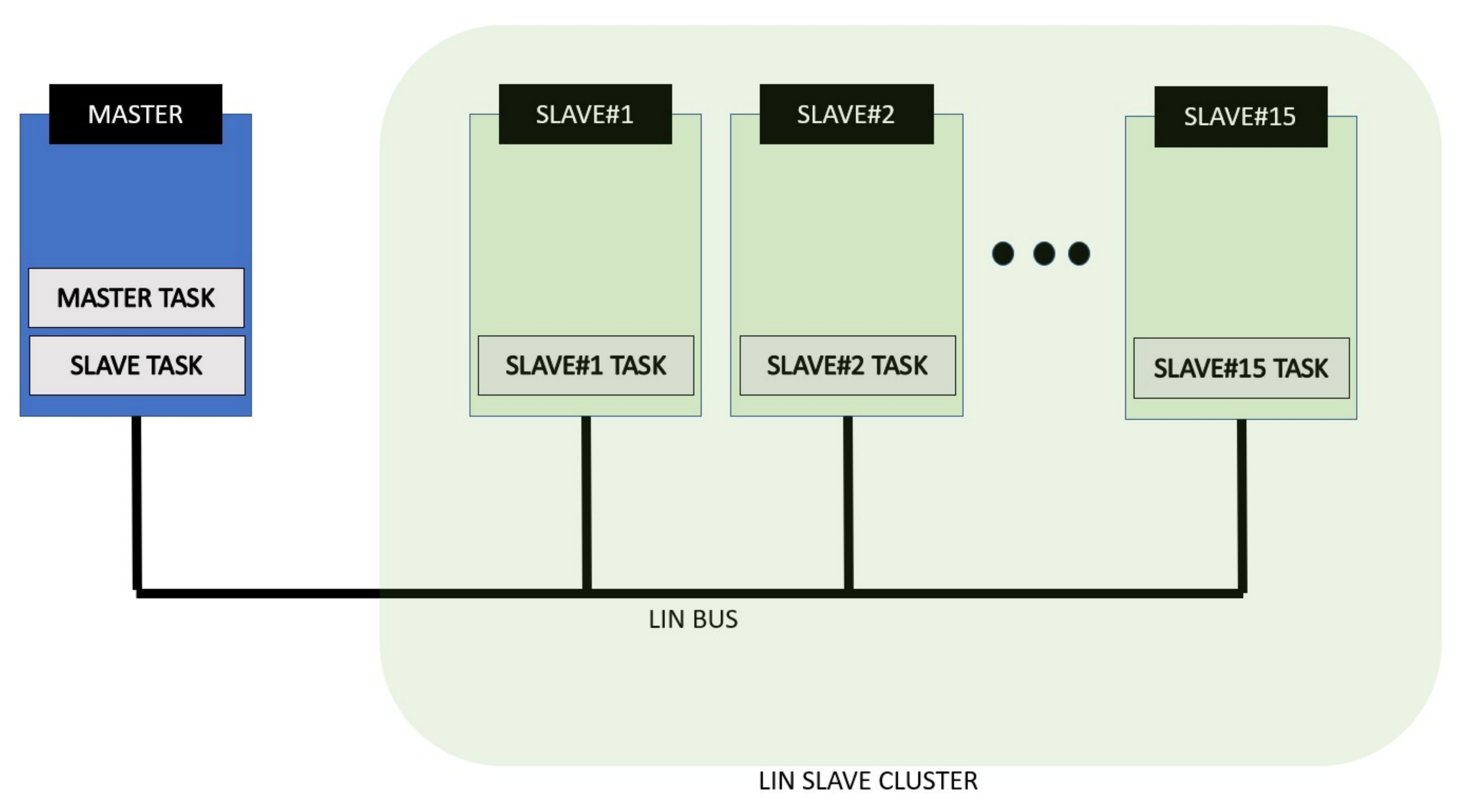}
    \caption{LIN Network Scheme}
    \label{fig:1}
 \end{figure}

The LIN is a broadcast network with serial master-slave communication and 16 nodes connectable to the bus. A single master manages up to 15 slaves for each LIN network (\autoref{fig:1}).

The main advantages of the LIN are the simplified wiring (i.e., a single wire supported by the major manufacturers), no licenses, and self-synchronization. These factors make LIN very competitive in terms of costs. The limitations are a slow baud rate up to 20K bit/s and a small 8-byte data frame. The master manages the LIN network. It starts the communication with the slaves. A slave contacted by the master replies with a data message. In this scheme, the master polls each slave on a time base. The master assigns a time slot for the slave to reply. This approach makes collision detection unnecessary on a LIN network. This means that only one device, the master, requires a precise oscillator.

The LIN nodes are based on microcontrollers. 
The LIN is compatible with the Universal Asynchronous Receiver Transmitter (UART) and the Serial Communications Interface (SCI) specifications, making hardware and software implementations simple.
  
\begin{figure*}[hbt]
    \centering
    \includegraphics[width=0.7\textwidth]{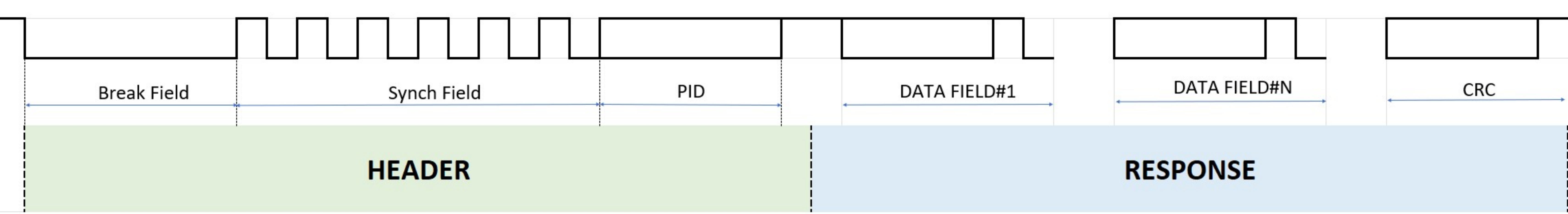}
    \caption{LIN Frame Format}
    \label{fig:2}
 \end{figure*}

The LIN single wire bus can reach up to 19.2 kbit/s, with a maximum length of 40 meters, although the LIN 2.2 specifications have increased the communication speed up to 20 kbit/s. The LIN master and slave frames are different (\autoref{fig:2}). The master frame comprises three fields: break field, synch field, and a protected identifier (PID), i.e., a 6-bit field identifying the target slave. The slave data frame includes up to 8 data bytes followed by an 8 bits Cycling Redundancy Check (CRC).

The battery voltage provides the nominal operating voltage for the LIN bus. The sender and receiver have different voltage level requirements. For the dominant bit (logic 0), the sender forces a voltage up to 20\% of the battery level on the LIN bus. The receiver interprets a dominant bit when it reads a voltage on the bus lower than 40\% of the battery level. For a recessive bit (logic 1), the sender applies 80\% of the battery voltage on the bus. At the same time, the receiver interprets a recessive bit reading a level higher than 60\% of the reference battery voltage. Different voltage thresholds between sender and receiver handle the possibility of groud shiftings in a vehicle bus, making the entire system more robust to occasional voltage dropping.


 \begin{figure}[htb]
   \centering
    \includegraphics[width=\columnwidth]{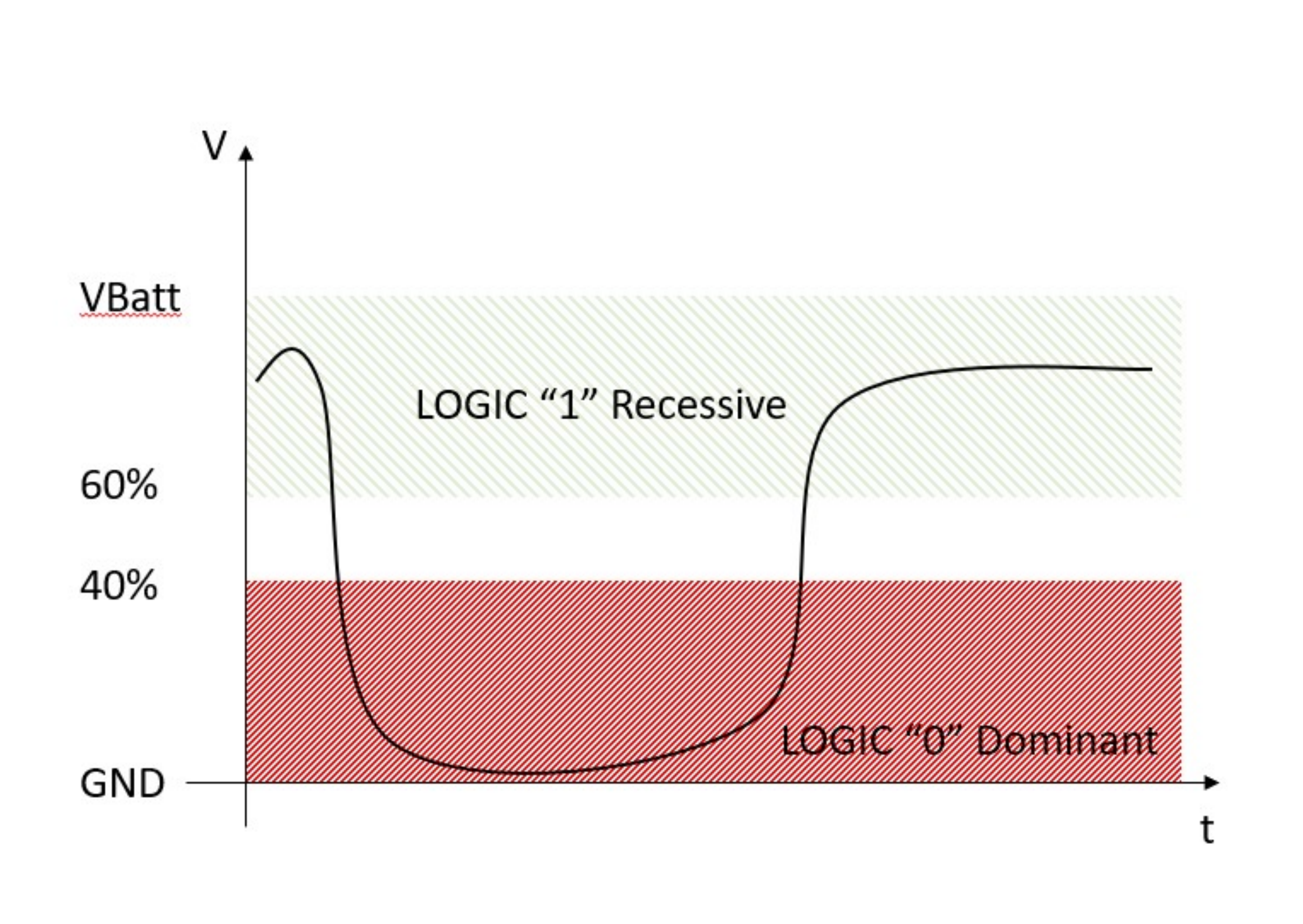}
    \caption{LIN Electrical Signal}
    \label{fig:signal}
 \end{figure}
 
Typical applications of the LIN in automotive include cruise windshield wipers, turning signals, climate sensors, mirrors, door locks, seat motors, and sensors and actuators belonging to the Powertrain perimeter like the Mass Air Flow (MAF) sensor and cooling fans.

\section{LIN attack vector analysis}
\label{sec:attack}

ECUs act as masters in a LIN network, while sensors and actuators are usually slave devices. Every LIN transaction follows the same schema. The master sends a header that includes a PID identifying a task carried out by a specific slave node (e.g., ask a sensor to report the measured physical quantity or command an actuator to set itself to a new target position). 

LIN slaves are the primary attack vector against LIN networks due to their high exposure to vulnerabilities. 
The literature reports four feasible primary attacks able to compromise the security of the LIN bus \cite{9465340,articleST,ELREWINI2020100214}.

In the \emph{message spoofing attack}, the attacker 
sniffs the bus traffic to identify the proper time slot to inject a spoofed message directed to a victim node.
Spoofed messages can be used to put to sleep a slave node, alter the SYNC field to tamper with synchronization, or inject illegitimate messages. In general, this attack aims to destroy the bus communication. Technical ability is not required to mount this attack. The attacker can exploit the dominant and recessive electrical states to cause frame corruption and Denial of Service (DoS).

\begin{figure}[htb]
   \centering
    \includegraphics[width=\columnwidth]{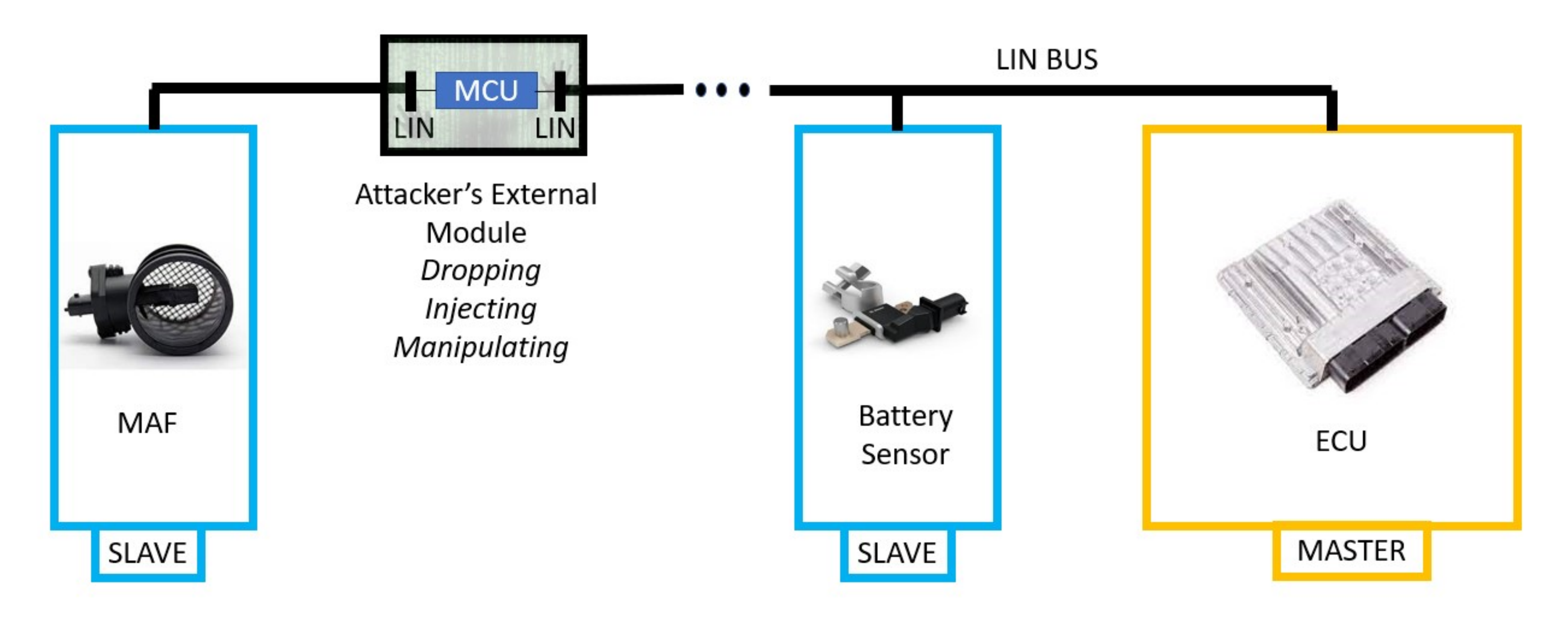}
    \caption{LIN MitM Attack Scheme}
    \label{fig:5}
 \end{figure}

Similar to the spoofing attack, the \emph{Man in the Middle attack (MitM)}  exploits an external malicious module to dissect a portion of the LIN bus. The malicious module can disconnect a victim node and a part of the LIN bus. Therefore, the attacker can hijack LIN frames from and to the disconnected LIN branch (\autoref{fig:5}). 
 
\begin{figure}[htb]
   \centering
    \includegraphics[width=0.8\columnwidth]{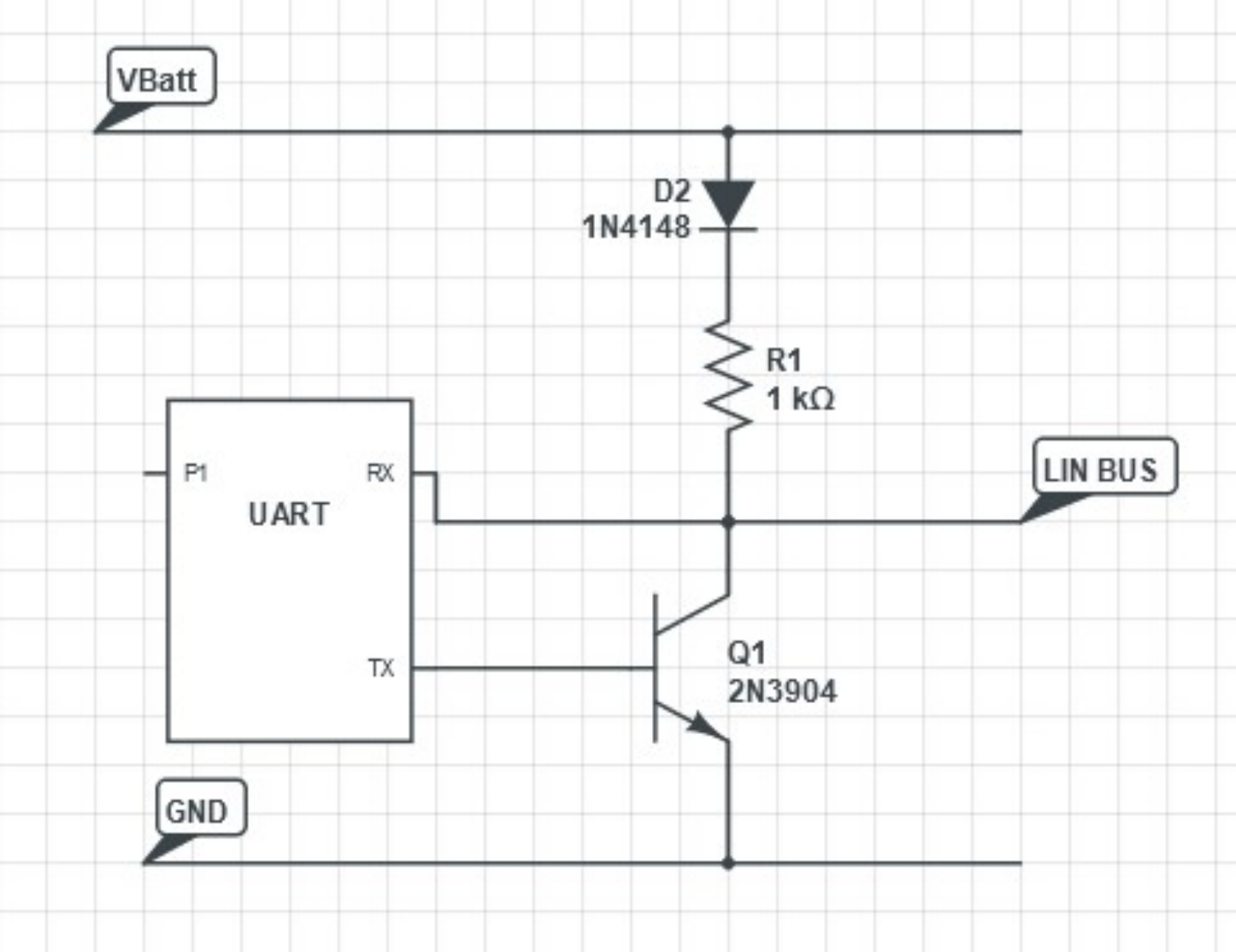}
    \caption{LIN Transceiver Scheme}
    \label{fig:uart}
 \end{figure}

The \emph{response collision attacks} happens when an illegitimate response message is transmitted together with a legitimate frame. According to the specifications, if a node detects a collision, it asserts an error bit and stops transmitting, waiting for the next transmission slot. A collision is detected when, during transmission, the electrical level observed on the bus is not coherent with the transmitted logical level. Checking the electrical level of the bus during transmission is possible thanks to the electrical implementation of the LIN transceiver (\autoref{fig:uart}).

The attacker can either send a false header playing as a master node or wait for the legitimate master frame and manipulate the message \cite{articleST}. The goal is to trigger an unexpected response colliding with the legitimate one. The legitimate slave node drops the transmission while the attacker can send an illegitimate response frame to the master node. The master node considers the attacker's message legit when the checksum code is correct. 

The \emph{header collision attack} is similar to the response collision attack, but the attack vector is now a master header frame instead of a slave response frame. The attacker injects an illegitimate header on the LIN bus to provoke collision with an allowed header. Again, in case of collision, the master stops transmitting, and the attacker can inject a malicious frame that can redirect a request to a different slave. In this way, the attacker can tamper with the sequence of responses of the LIN network and even isolate a victim node. This kind of attack can slide down a vehicle's windows, lock or unlock the car, or lock steering wheels while vehicles are traveling along the road.

The main difficulty of injecting an incorrect response on the LIN network lies in exploiting physical access to the LIN bus. Direct access through external modules, as shown in \autoref{fig:5} provides complete control of the bus. However, it is possible to gain partial LIN access through the CAN network. Usually, the master LIN nodes are connected to the CAN bus \cite{DENG2017131}. An attacker can reach the LIN bus through the CAN On-Board Diagnostic (OBD) port mounted in a vehicle cabin, using recent hacking techniques \cite{MARTINEZCRUZ20211,hackLIN}. 

To avoid the response and header collision attacks, Takahashi et al. \cite{articleST} suggest that a slave node sends out an abnormal signal, which would overwrite a false message sent by an attacker if it detects that the bus value does not match its response. The solution is limited because the slave communicates when the master releases the appropriate time slot, so the countermeasure is not immediate. Additional suggestions include incorporating MAC and assigning essential data to the first byte of transmission, as the first byte is more difficult to corrupt. The main limitation is that the MAC code erodes the frame data payload. The LIN protocol has a maximum data payload of 8 bytes. By NIST's guidelines \cite{CMAC}, a robust MAC code cannot be less than 64 bits, which is precisely the entire LIN payload length. The current mitigation is to reduce the MAC digest to 4 bytes providing protection that does not meet the automotive security standards.


\section{LIN-MM}
\label{sec:LINMM}

Multiplexed Message Authentication Code for Local Interconnect Network (LIN-MM) is a new solution to introduce message authentication code compliant with NIST's guidelines \cite{CMAC} in a standard LIN network without reducing the LIN payload size. LIN-MM is not intrusive and ensures full back compatibility with standard LIN devices. Moreover, LIN-MM eliminates the MAC transmission latency by transmitting it concurrently with the data payload. 

\begin{figure}[htb]
   \centering
    \includegraphics[width=\columnwidth]{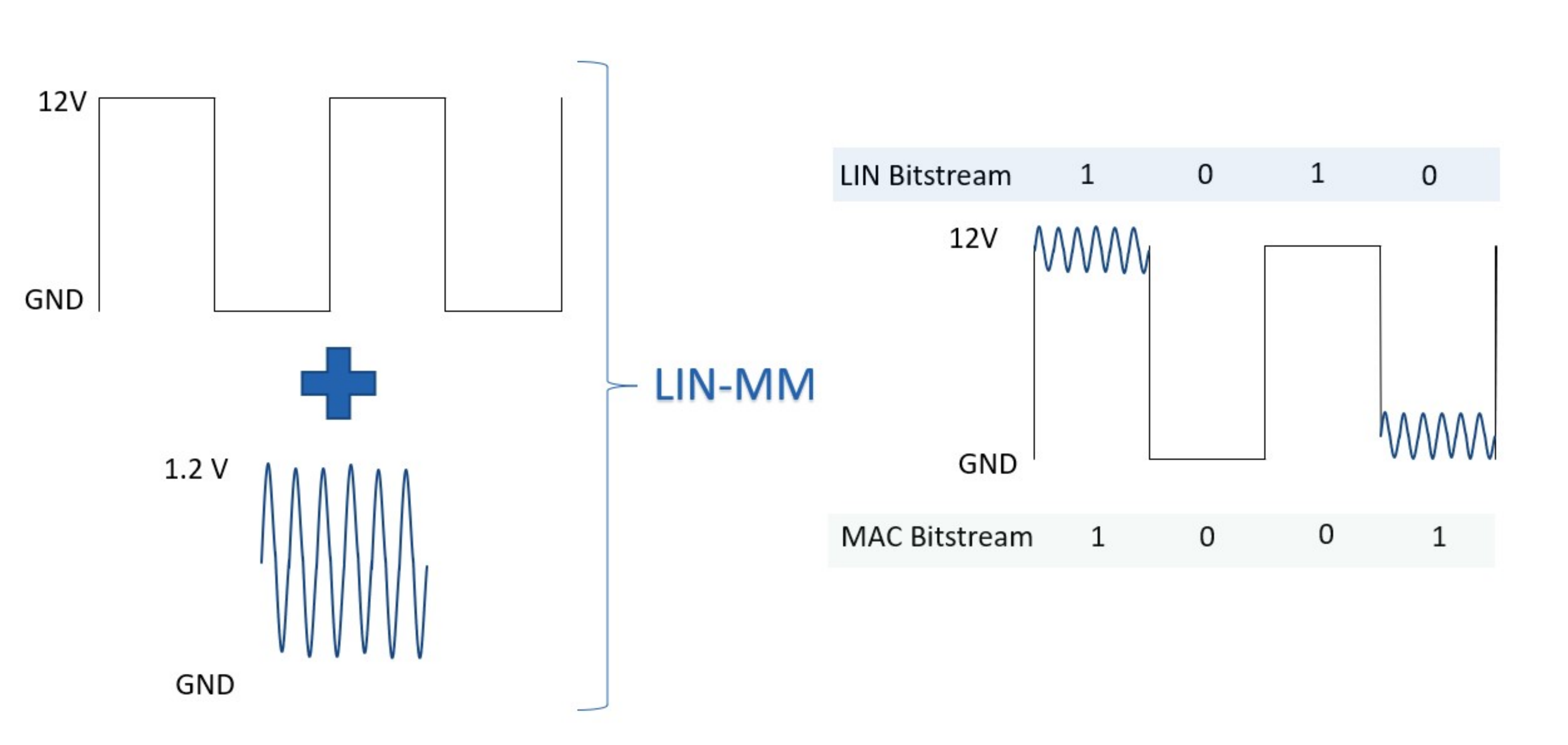}
    \caption{LIN-MM Physical Electrical Signal}
    \label{fig:6}
 \end{figure}

LIN-MM applies signal modulation to multiplex the MAC bitstream with the electrical signal of the LIN frame (\autoref{fig:6}). In particular, LIN-MM exploits On-Off Keying (OOK) \cite{onoffkeing}, one of the most straightforward digital modulation schemes, to transform the MAC bitstream into an electrical signal. OOK uses a carrier signal; it turns the carrier ``On'' when transmitting a logic `1' and turns it ``Off'' when transmitting a logic `0'. The modulated carrier is added to the original LIN electrical signal to multiplex the MAC bitstream with the LIN bitstream. The full 8-byte payload of a LIN response is available to implement advanced features using LIN-MM.

In its current implementation, LIN-MM is limited to the slave responses. The 8-byte payload of a slave response is enough to carry the information of a 64-bit MAC code compliant with NIST recommendations. Even if the same technique can be applied to the master header frames, the limited length only allows multiplexing short and less secure codes.

\subsection{LIN-MM Slave Architecture}

\begin{figure}[htb]
   \centering
    \includegraphics[width=\columnwidth]{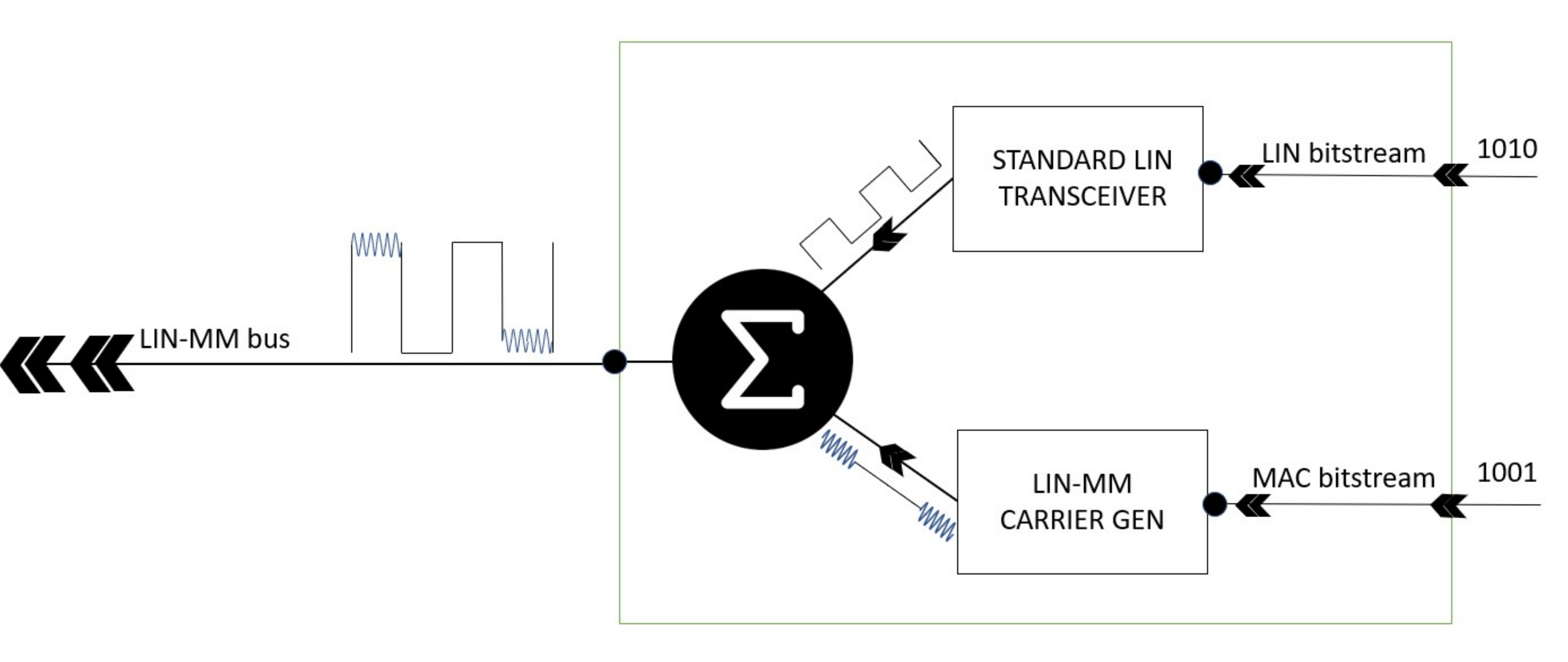}
    \caption{LIN-MM Slave Block-Scheme in trasmission}
    \label{fig:8}
 \end{figure}

As illustrated in \autoref{fig:8}, the LIN-MM Slave pairs a standard LIN transceiver with a custom LIN-MM carrier generator. This component generates a modulated carrier signal based on the MAC bitstream. The original LIN signal and the carrier signal are added to create the LIN-MM electrical signal. 

\begin{figure}[htb]
   \centering
    \includegraphics[width=\columnwidth]{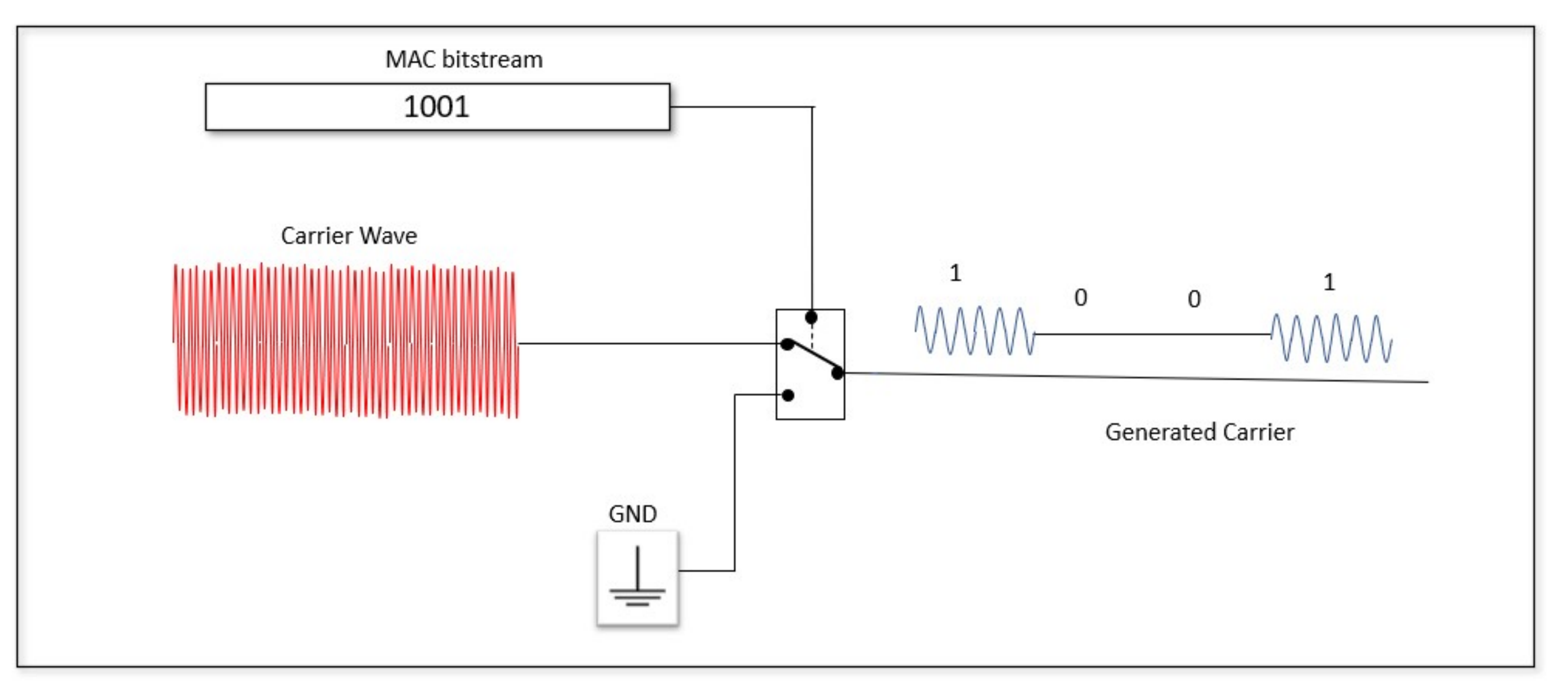}
    \caption{LIN-MM Carrier Generator Block-Scheme}
    \label{fig:9}
 \end{figure}

The LIN-MM carrier generator is composed of a multiplexer accepting two inputs: the carrier provided by a dedicated signal generator and GND (\autoref{fig:9}). The MAC bitstream controls the multiplexer commutation. It switches the multiplexer to the carrier signal when the corresponding MAC bit is '1'; it changes to GND otherwise. The standard LIN transceiver and the LIN-MM carrier generator are synchronized to multiplex the two signals properly.

The carrier frequency ($f_c$) must be selected to enable its isolation from the LIN signal through filtering even in case of noise. The proposed implementation exploits a 100 kHz sinusoidal signal that can be easily separated from the 20 kbit/s LIN signal. The carrier frequency is synchronous with the pure LIN signal, and both share the same sample time windows. Considering  $V_{batt}=12V$, the carrier amplitude is set to 1.2V to keep the generated signal within the LIN electrical specifications (\autoref{sec:LINMM}). 

\subsection{LIN-MM Master}

\begin{figure}[htb]
   \centering
    \includegraphics[width=\columnwidth]{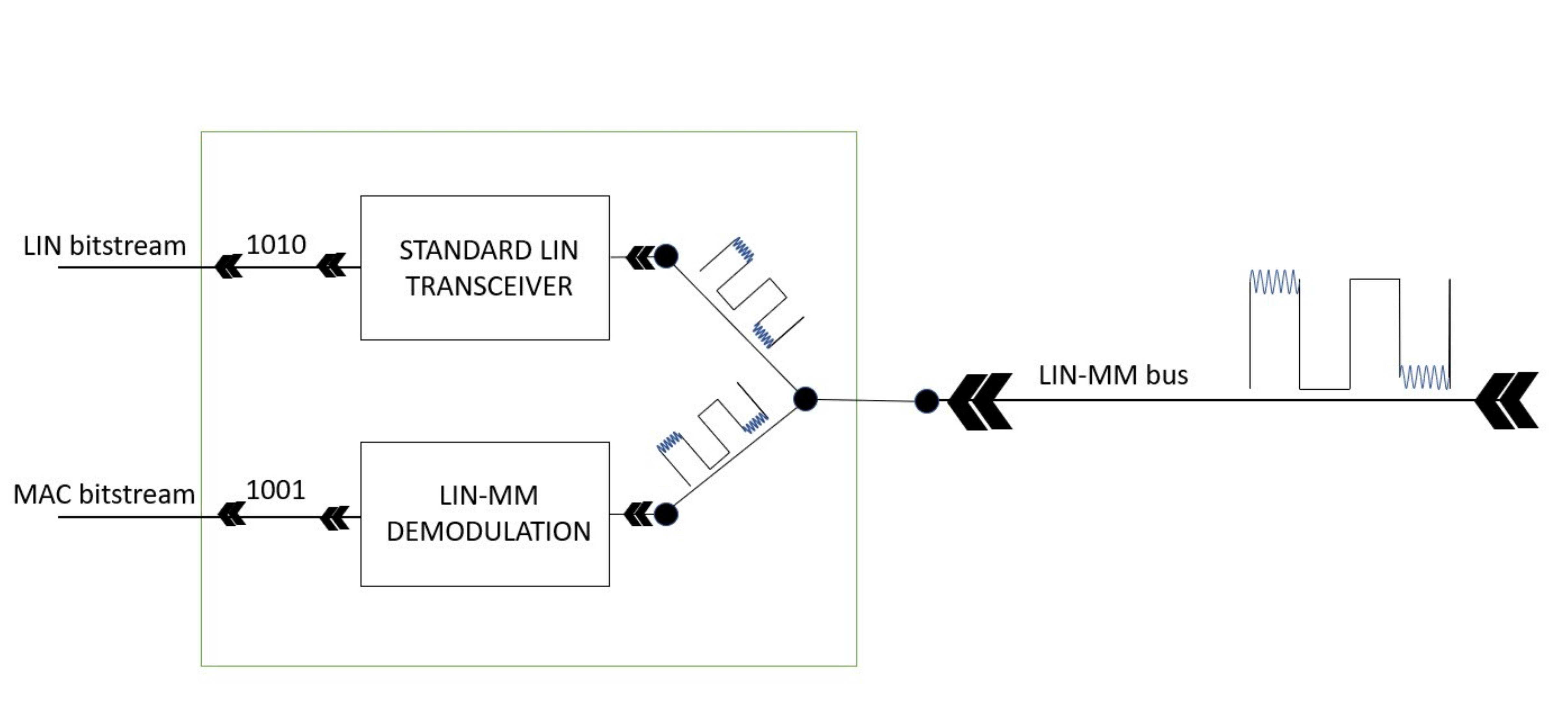}
    \caption{LIN-MM Master Block-Scheme in receiving}
    \label{fig:10}
 \end{figure}

\autoref{fig:10} shows the architecture of the LIN-MM Master composed of two main blocks: a standard LIN transceiver and a demodulation block. The standard LIN transceiver processes the LIN-MM signal. The MAC multiplexed signal is noise canceled by the internal transceiver filters. The demodulation block is introduced to demodulate and reconstruct the MAC bitstream from the physically received signal.

\begin{figure}[htb]
     \centering
     \begin{subfigure}[b]{0.48\textwidth}
         \centering
    	 \includegraphics[width=\columnwidth]{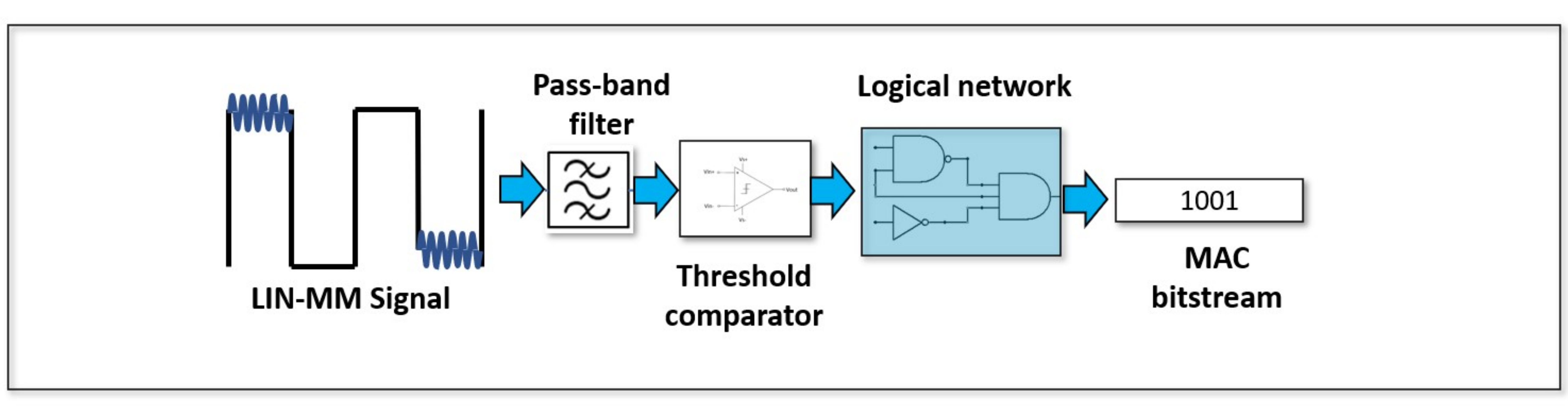}
    	 \caption{LIN-MM Demodulation Block pipeline}
         \label{fig:11}
     \end{subfigure}
     \hfill
     \begin{subfigure}[b]{0.48\textwidth}
         \centering
    	 \includegraphics[width=\columnwidth]{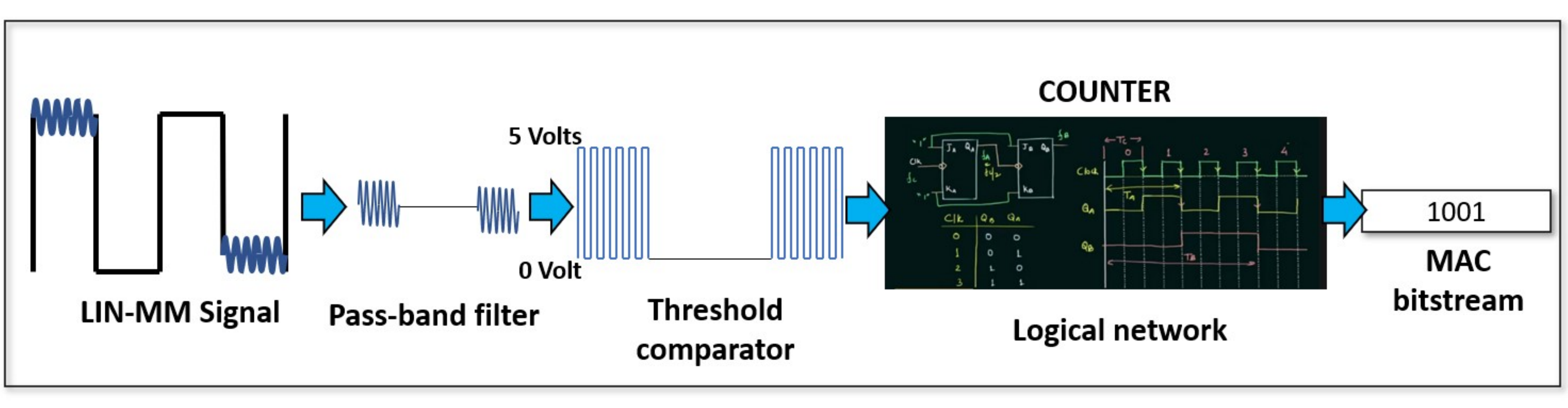}
   	 \caption{LIN-MM Demodulation Block Signal Profile}
   	 \label{fig:asssigprof}
    \end{subfigure}
        \caption{LIN-MM Demodulation Block Scheme }
       \label{fig:LINMMSlavesignal}
\end{figure}

The LIN-MM demodulation block comprises three subsystems (\autoref{fig:11}). \autoref{fig:LINMMSlavesignal} shows the effect of these subsystems on the processed signal. 

A pass-band filter with a center frequency $f_c$ at the carrier’s frequency isolates the carrier’s contribution from the rest of the signal. With $f_c=100kHz$ the filter is designed with a bandwidth $BW=50kHz$ corresponding to a low-pass frequency $f_l=75kHz$ and a high-pass frequency  $f_h=125kHz$.

A threshold comparator transforms the analog sinusoidal carrier into a signal in the digital domain with a bit transmission rate equal to the carrier frequency. 

Finally, a digital network reconstructs the MAC bitstream. It acts as a digital counter synchronized with the LIN bit signal. The counter is reset at the beginning of each LIN period (i.e., LIN bit transmission period) and counts the number of digital pulses generated by the comparator. At the end of the LIN period, a MAC logic '0' is reconstructed if the counter detected less than three pulses, a logic '1' otherwise. This approach allows for reliable MAC bitstream reconstruction, with a robust resilience to noise that may generate spurious spikes pulses. With this schema, the MAC bitstream can be reconstructed with a delay of a single LIN period.

 
\section{Experimental results}
\label{sec:results}

\begin{figure}[htb]
\centering
\includegraphics[width=\columnwidth]{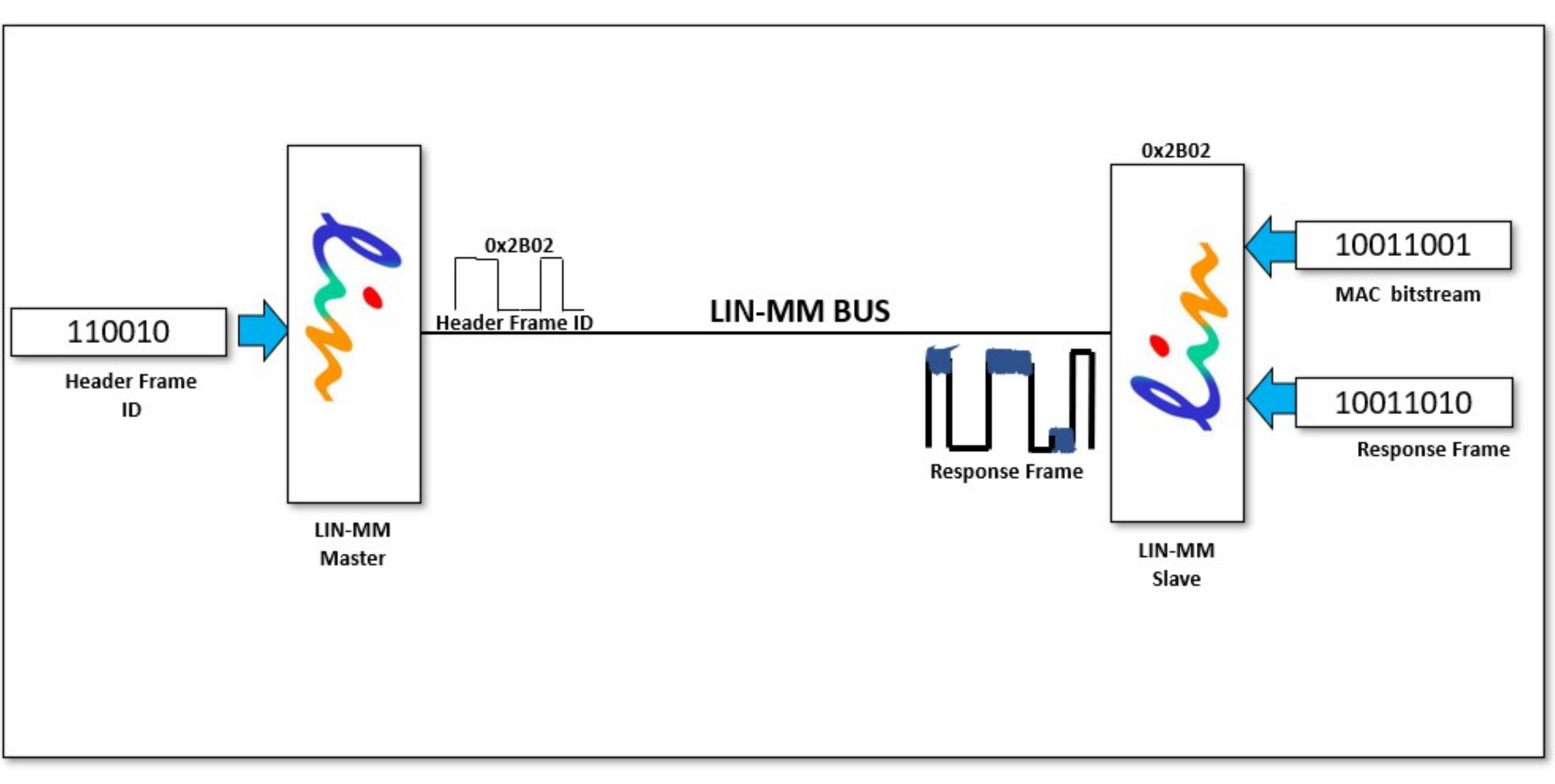}
\caption{LIN-MM Spice Model Block Scheme}
\label{fig:14}
\end{figure}

An LTSpice~\cite{ltspice} prototype implementation of a LIN-MM architecture composed of a Master and a Slave node was implemented to prove the feasibility of the proposed schema and therefore validate and verify the LIN-MM functionality.  \autoref{fig:14} shows a high-level block scheme of the prototype architecture. The master node sends a LIN header frame to the slave node that replies with a LIN response message. The response message encapsulates a 64-bit MAC digest for authenticating the transmitted data. The system works with a 19.2k bit/s baud rate. The entire architecture is implemented with a standard LIN transceiver.

\subsection{Functional validation}

\begin{figure}[htb]
\centering
\includegraphics[width=\columnwidth]{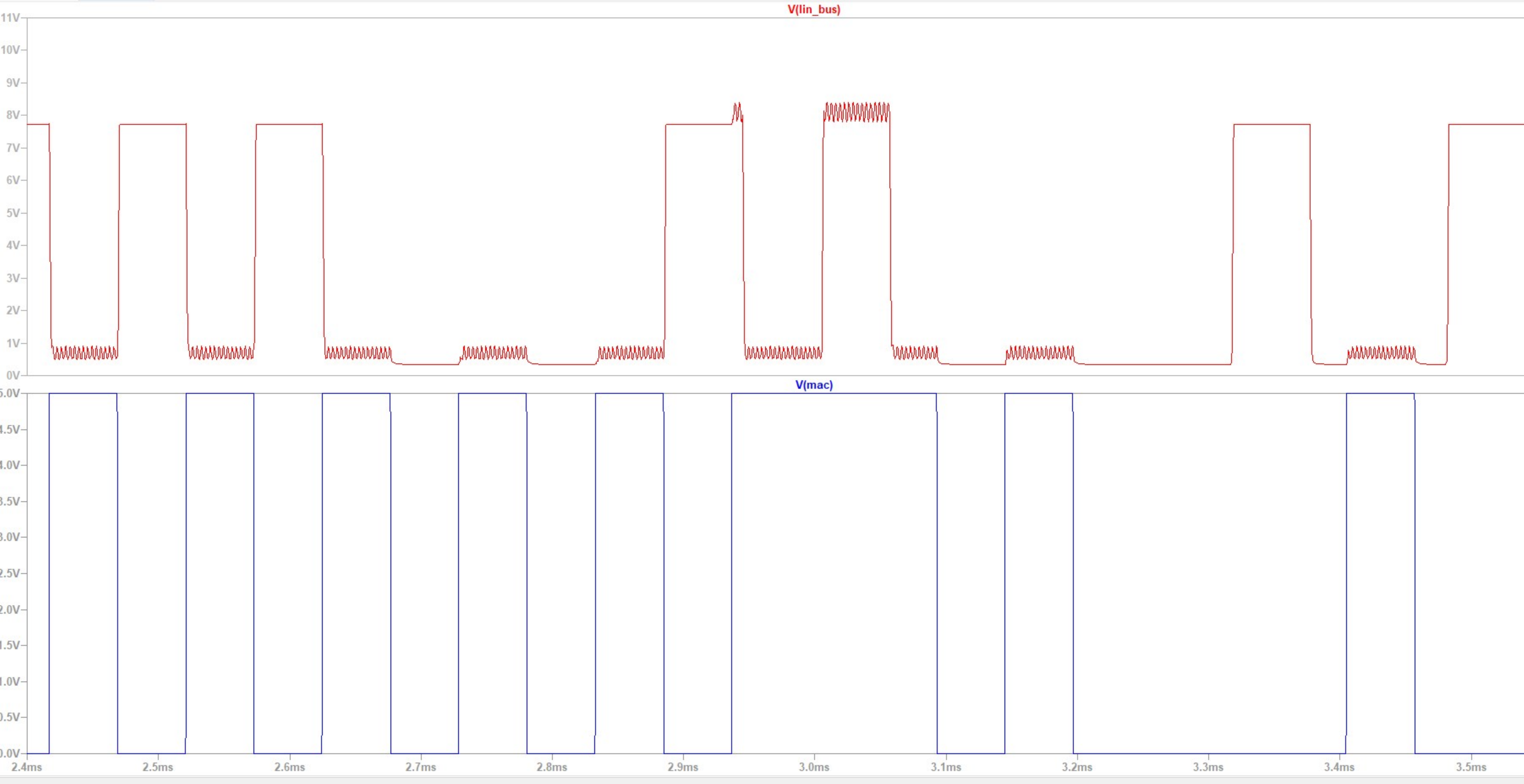}
\caption{LIN-MM Response Frame}
\label{fig:15}
\end{figure}

\autoref{fig:15} provides a functional validation of the proposed architecture showing the LIN-MM Response Frame, produced by the LIN slave node (\autoref{fig:14}). The first signal (red) shows the physical, electrical signal on the LIN bus. The second signal (blue) shows the MAC bitstream encapsulated in the LIN-MM signal.

As expected, the LIN-MM physical signal shows the presence of the carrier in correspondence to MAC bits at logic  '1'. In contrast, no carrier is present in correspondence of MAC bits at logic '0'. 

\begin{figure}[ht!]
\centering
\includegraphics[width=8cm]{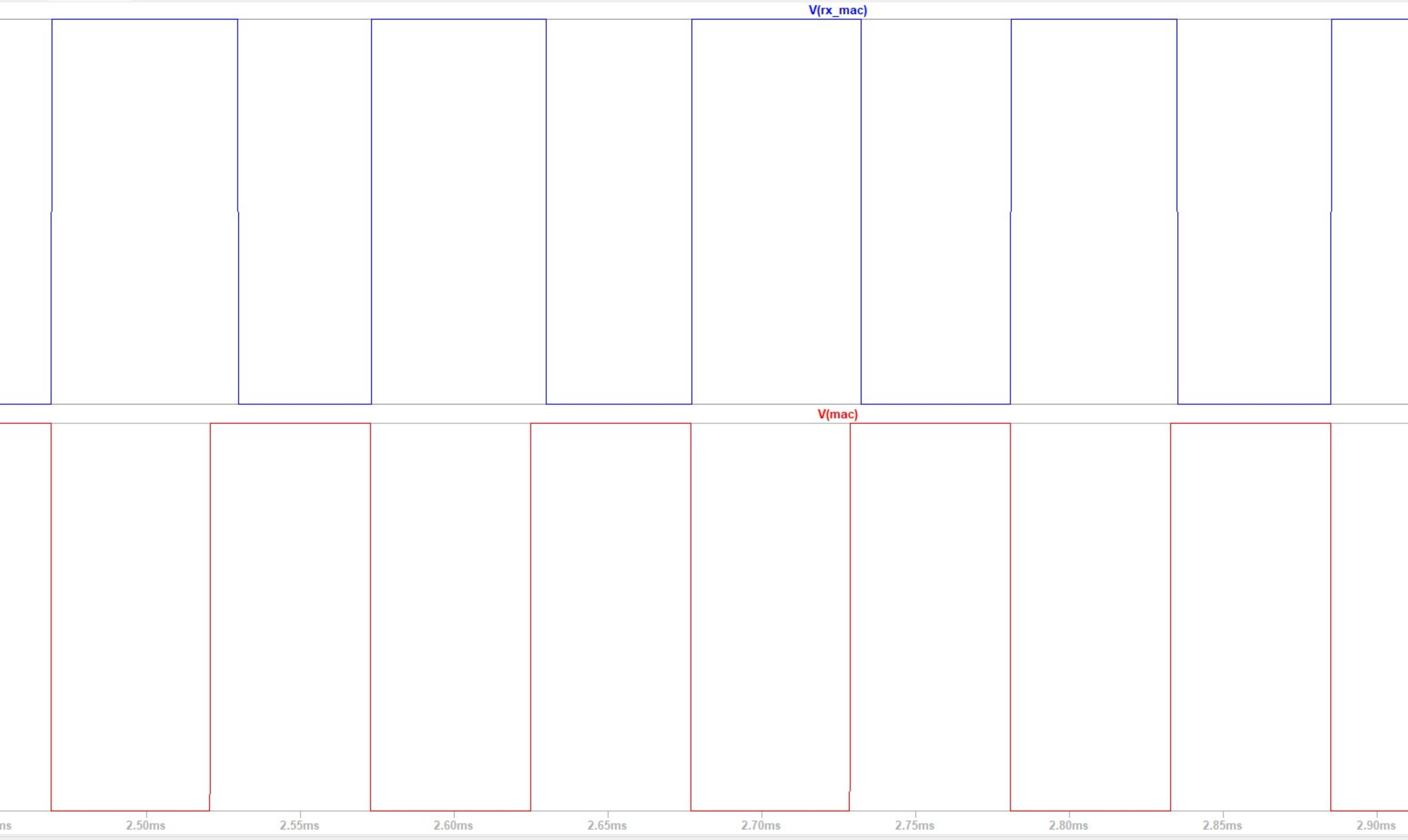}
\caption{MAC bitstream time propagation.}
\label{fig:18}
\end{figure}

\autoref{fig:18} shows the MAC bitstream propagation time. The red signal is the MAC bitstream generated by the LIN-MM slave, while the blue line is the MAC bitstream reconstructed by the LIN-MM Master. The MAC bitstream propagation latency time is limited to one LIN period. 
 
\subsection{Overhead evaluation}

Implementing the LIN-MM requires introducing an additional plug-in module to all nodes that require secure LIN communication.  

Comparing the standard LIN transceiver hardware with the LIN-MM transceiver, we estimate a 2\% cost overhead. This estimation is obtained by using the cost of a standard LIN transceiver as a reference and considering the cost of the additional discrete components required to implement the LIN-MM modulation and demodulation blocks. This overhead is acceptable considering that the only alternative solution to guarantee adequate security levels would be to migrate from the LIN protocol to a more secure CAN protocol that would introduce definitively higher costs.  

The LIN-MM receiver obtains the full MAC bitstream 50us after receiving the LIN payload with a very short latency time. LIN-MM introduces a significant performance improvement compared to today's insecure implementations with a 32-bit MAC code embedded in the payload, where the latency time is around 190ms.  

Eventually, let us discuss the LIN-MM impact on the hardware and vehicle architecture. The LIN-MM solution is fully compatible with standard hardware and can work on a hybrid network. It permits upgrading the sensible asset nodes to LIN-MM while keeping nodes with no security implications on legacy hardware.  

In conclusion, the experimental results support the initial expectations. The LIN-MM concept achieves the main features to multiplex the MAC code directly at the LIN physical signal level. This achievement is crucial to confirm the security level improvement of LIN-MM compared with the standard LIN network.

\subsection{Security analysis}

The keystone of LIN-MM is a mechanism to multiplex at the electrical level a MAC digest with a standard LIN signal. 

The proof of the security of the LIN-MM solution holds under the infeasibility hypothesis. We assume the use of state-of-the-art secure cryptographic algorithms with proper key lengths. 

Employing a state-of-the-art automotive hardware control module, LIN-MM can be implemented to work with a state-of-the-art Cipher-based Message Authentication Code (CMAC) based on the Advanced Encryption Standard 128-bit Cipher (AES128) with Block Chaining (CBC) modality. The introduction of the MAC code guarantees integrity and authentication on LIN communication with significant mitigation of the attacks introduced in \autoref{sec:attack}. CMAC-128bits with a truncated digest to 8 bytes guarantees a considerable security level increment. Sixty-four bits are the minimum digest’s length for considering the MAC code as secure. Today, solutions embedding a 32-bit digest in the LIN data payload cannot guarantee this level of security. 

By guaranteeing integrity and authenticity of LIN response frames,  LIN-MM ensures resiliency from the attacks that exploit the response frame as a vector, i.e., Spoofing, Man in the Middle (MitM), and Response collision attacks. 

On the contrary, LIN-MM is not a viable security mechanism for those attacks that use the header frame as a vector, i.e., Header collision attacks. The primary constraint, in this case, is that the primary constraint is enough to multiplex a robust MAC code. Moreover, the LIN-MM is not effective as a countermeasure for \emph{Denail of Service (DoS)} attacks.  

The LIN-MM architecture requires sharing of a secret key between the master and the slave nodes. Storing this secret key is not a significant issue, given that all LIN nodes are based on microcontrollers. In this context, the key management infrastructure becomes an important issue that must be considered before deploying this solution at a commercial level.

\section{Conclusion}
\label{sec:conclusions}

LIN-MM is an attempt to upgrade the standard adopted LIN hardware in the automotive domain to fulfill the new challenges of the upcoming security regulations and requirements, keeping the associated cost as low as possible. LIN-MM is back-compatible with today’s LIN devices. It can work in a hybrid network that mixes LIN and LIN-MM devices, making it a very flexible solution. This approach reduces the costs of upgrading the existing vehicle fleet to cyber-secure vehicle regulations.

The conducted security analysis shows a visible improvement in terms of security for the LIN network, especially on what concerns the first three kinds of attack described in \autoref{sec:attack}. The last category of attacks, i.e., Header collision attacks, is still in place with no mitigation driven by LIN-MM architecture.

Currently, we are working to enhance the LIN-MM architecture on a new version preliminary named (Type-B) to address this vulnerability by providing integrity to the LIN Header Frame.

\nocite{*}
\bibliographystyle{IEEEtran}
\bibliography{LINMM_MAIN}

\end{document}